\documentclass[11pt]{article}

\usepackage{amssymb,amsmath}

\newcommand{\newsection}{\addtocounter{section}{1}
\setcounter{equation}{0}\section}

\newcommand{\mbf}[1]{{\boldsymbol {#1} }}

\newcommand{\complex}{{\mathbb C}} 
\newcommand{\zed}{{\mathbb Z}} 
\newcommand{\real}{{\mathbb R}} 
\newcommand{\zeds}{{\mathbb Z}} 
\newcommand{\mat}{{\mathbb M}} 
\def\module{{\cal E}}
\newcommand{\id}{{1\!\!1}} 
\def\alg{{\cal A}}
\def\heisen{{\cal L}}

\def\vp{{\mbf p}}
\def\vq{{\mbf q}}
\def\vn{{\mbf n}}
\def\vm{{\mbf m}}

\def\vnu{{\mbf \nu}}
\def\torus{{\mathbf T}}
\def\sphere{{\mathbf S}}
\def\proj{{\mathsf P}}

\def\nn{\nonumber}
\newcommand{\tr}[1]{\:{\rm tr}\,#1}
\newcommand{\Tr}[1]{\:{\rm Tr}\,#1}
\def\e{{\,\rm e}\,}

\newcommand{\non}{\nonumber \\}

\def\beq{\begin{equation}}
\def\eeq{\end{equation}}
\def\bea{\begin{eqnarray}}
\def\eea{\end{eqnarray}}
\def\bd{\begin{displaymath}}
\def\ed{\end{displaymath}}

\def\dd{{\rm d}}

\def\ii{{\,{\rm i}\,}}

\renewcommand{\r}{\prime}

\begin{document}
\begin{flushright}

\baselineskip=12pt
MCTP--03--09\\ HWM--03--4\\ EMPG--03--05\\ hep-th/0302195\\
\hfill{ }\\ February 2003
\end{flushright}

\begin{center}

\baselineskip=24pt

{\Large\bf LECTURES ON TWO-DIMENSIONAL\\ NONCOMMUTATIVE GAUGE
THEORY
  \\ 1. Classical Aspects}
\footnote{\baselineskip=12pt Based on invited
  lectures given by the second author at the {\it 2nd Summer School in
  Modern Mathematical Physics}, September 1--12 2002, Kopaonik,
  Yugoslavia; at the {\it European IHP Network Conference on Quantum
  Gravity and Random Geometry}, September~7--15 2002, Orthodox
  Academy of Crete, Kolympari, Greece; and at the {\it International
  Workshop on Quantum Field Theory and Noncommutative Geometry},
  November~26--30 2002, Tohoku University, Sendai, Japan. To be
  published in the School proceedings by Belgrade Institute of Physics.}

\baselineskip=12pt

\vspace{10mm}

{\bf L.D. Paniak}
\\[2mm]
{\it Michigan Center for Theoretical Physics, University of
Michigan\\ Ann Arbor, Michigan 48109-1120, U.S.A.}\\  {\tt
paniak@umich.edu}
\\[4mm]

{\bf R.J. Szabo}
\\[2mm]
{\it Department of Mathematics, Heriot-Watt University\\ Scott
Russell
  Building, Riccarton, Edinburgh EH14 4AS, U.K.}\\  {\tt
R.J.Szabo@ma.hw.ac.uk}
\\[10mm]

\begin{minipage}{11cm}
\small \baselineskip=12pt

These notes comprise the first of two articles devoted to the
construction of exact solutions of noncommutative gauge theory in
two spacetime dimensions. This first part deals solely with the
classical theory on a noncommutative torus. Topics covered include
a mathematical introduction to the geometry of the noncommutative
torus, the definition, properties and symmetries of noncommutative
Yang-Mills theory, and the complete solution of the classical
field equations.

\end{minipage}

\end{center}

\vspace{5mm}

\baselineskip=12pt

\newsection*{\sc 1.~Introduction}

In these lecture notes we will describe how to explicitly
construct exact, analytic solutions of noncommutative gauge theory
in two dimensions. The analysis is naturally divided into two
articles. In this first part the classical field theory will be
studied in some detail. This is done in order to provide a
pedagogical introduction to the relevant mathematical concepts
which are presently at the forefront of the field. As such, most
of the analysis in the following consists of material that has
been known to mathematicians for well over ten years now. The
second part of these lecture notes will analyse the quantum field
theory in detail and will deal with more recent results. As we
will see there, the quantum theory is semi-classically exact and
hence is determined in large part by the classical aspects dealt
with here.

\subsection*{\sc 1.1.~Some Background and Motivation}

Yang-Mills theory on the noncommutative torus was first studied in
the mathematics literature in the late 1980's at the classical
level~\cite{cr,rieffel}.
It represents one of the few instances in which a physical
model has attained popularity among physicists, yet was first
introduced and studied by mathematicians. Although field theories
on noncommutative spaces have been widely studied in the past in a
variety of different contexts, the recent surge of interest in the
subject has primarily arisen from the fact that these models
naturally arise in string theory \cite{cmrMatrix,sw}. When a D-brane
is placed in the background of certain non-vanishing supergravity fields, the
dynamics of some low-energy excitations of the open strings which
attach to them are described by a noncommutative field theory. As
field theories, these models retain much of the non-locality which
is present in string theory. It is hoped that noncommutative field
theories will reveal many of the general features of string theory
but within the simpler setting of quantum field theory. Recent reviews
of the subject within this context, along with exhaustive lists of
references, may be found in~\cite{ks}--\cite{sz1}.

Besides being of interest from the point of view of string theory,
noncommutative field theories challenge the conventional wisdom of
ordinary quantum field theory. They are non-local and contain
infinitely many derivative interactions which would appear to lead
to non-renormalizability problems in a full quantum theory. There
are indications though that these issues are dealt with by the
specific structure of higher-derivative interactions introduced by
noncommutative geometry. Unfortunately, answering such questions
for a noncommutative field theory is complicated by the mixing of
low and high momentum modes in loop diagrams which ruins the
conventional Wilsonian renormalization scheme that requires a
distinct separation of energy scales. This effect is commonly
called ``UV/IR mixing''. It appears to make the renormalization of
these theories a complete disaster, but it is not yet well
understood if these effects are really artifacts of the way we are
treating these models using perturbation theory, or if they also
persist at a full non-perturbative level.

The non-locality of noncommutative field theories leads to many
interesting phenomena which makes these models interesting in
their own right as potentially well-defined, non-local quantum
field theories. Furthermore, because of ``Morita duality'', it has
even been suggested that noncommutative gauge theories may shed
new light on non-perturbative aspects of {\it ordinary} Yang-Mills
theory. Another exotic feature of  the quantum theory is the
property that the perturbation series does not reduce smoothly to
its commutative version, but rather exhibits poles in the
noncommutativity parameter. Again, this ``$\theta$-smoothness'' behaviour may
simply be a perturbative artifact of the theory and can disappear
in the full quantum field theory.

These open questions motivate the search for models on
noncommutative spaces which are exactly solvable in order that we
may observe whether or not these unusual effects persist
non-perturbatively. Until very recently, surprisingly little
attention has been paid to finding such theories. In these lecture
notes we will describe and analyse one such field theory, namely
two-dimensional Yang-Mills theory on the noncommutative torus.
Various aspects of this model and its analogue on the noncommutative
plane have been the subject of recent
investigation~\cite{poly}--\cite{inprep2}.  In
the course of analysing this model we will exploit some of the
beautiful mathematics behind noncommutative geometry and
two-dimensional gauge theories. As such, these lecture notes also
provide to some extent an introduction to the building blocks of
noncommutative geometry. Even though we restrict our discussion to
the simplest noncommutative space, this case nevertheless exhibits
most of the rich features of noncommutative geometry. This
setting also enables us to precisely define and describe the
exotic properties of noncommutative gauge theories, and to show
how noncommutative geometry may be exploited to explicitly solve the
field theory.

\subsection*{\sc 1.2.~Outline}

The outline of material contained in these notes is as follows. In
section~2 we summarize all the relevant mathematical details from
noncommutative geometry that we shall need. A more general and
detailed presentation may be found in \cite{connesbook}. We present a concise
introduction to the geometry of the two-dimensional noncommutative
torus, the construction of differential structures and bundles
over it, and the corresponding notions of connection and
curvature. In section~3 we then define classical Yang-Mills theory
on the noncommutative torus and describe the rich geometrical
structure underpinning the gauge group in this case. We also
describe the integrability properties underlying the exact
solvability features of this model. In section~4 we explicitly
construct all solutions to the classical field equations of
noncommutative Yang-Mills theory. Finally, in section~5 we
introduce the important notion of Morita equivalence, discuss its
physical implications, and show that noncommutative gauge theory
is invariant under it.

\newsection*{\sc 2.~Background from Noncommutative Geometry}

This section comprises a quick introduction to the geometry of the
noncommutative torus. We shall be fairly mathematical here in
order to present the results in a precise and rigorous fashion.
Later on this level of formality will become a true asset in
arriving at the exact solution of the noncommutative field theory.

\subsection*{\sc 2.1.~The Noncommutative Torus}

We shall begin by introducing the noncommutative torus as an
abstract object and then below describe it from a more heuristic
level. Let $\alg=\alg_\theta$ be the noncommutative associative
$*$-algebra, with unit $\id$ and conjugation involution $\dag$,
generated by two elements $\hat Z_1$ and $\hat Z_2$ which obey the
relations \beq \hat Z_1\,\hat Z_2=\e^{2\pi\ii\theta}~\hat
Z_2\,\hat Z_1 \ , ~~ \hat Z_i^\dag=\hat Z_i^{-1} \ ,
\label{NCtorusrels}\end{equation} where the real number $\theta\in(0,1)$ is
called the noncommutativity parameter. We shall usually assume
that it is irrational-valued. The ``smooth'' completion of this
algebra consists of elements $\hat f$ which are formal power
series of the form \beq \hat
f=\sum_{\vm\in\zeds^2}f^{~}_{\vm}~\e^{\pi\ii\theta\,m_1m_2}~ \hat
Z_1^{m_1}\,\hat Z_2^{m_2} \ , \label{hatfpower}\end{equation} where
$\vm=(m_1,m_2)$. To ensure convergence of the expansion we assume
that $f_\vm$ are Schwartz sequences (of sufficiently rapid
decrease at $m_1,m_2\to\pm\,\infty$), and the $\theta$ dependent
phase factor is inserted for convenience to enforce the symmetric
ordering of all noncommuting products. The hat notation here is
used in order to distinguish these abstract algebraic objects from
their corresponding spacetime fields that will be introduced
shortly.

\subsection*{\sc 2.2.~Derivatives}

At this abstract level we can define derivatives as linear maps
$\hat\partial:\alg\to\alg\otimes\heisen^*$, where
$\heisen=\heisen_\phi$ is the Heisenberg Lie algebra with ``Planck
constant'' $\phi$. These maps define a Lie algebra homomorphism,
$[\hat\partial_X,\hat\partial_Y]=\hat\partial_{[X,Y]}~~\forall
X,Y\in\heisen$, which in components are defined by the commutation
relations \beq
\left[\hat\partial_1\,,\hat\partial_2\right]=\ii\phi \ , ~~
\left[\hat\partial_i\,,\,\hat Z_j\right]=2\pi\ii\,\delta_{ij}~\hat
Z_j \ . \label{hatdeldef}\end{equation} In the first commutator the real
number $\phi$ implicitly multiplies the unit element $\id$ of the
algebra $\alg$, which for brevity we will not indicate explicitly.
We extend $\hat\partial$ to the whole of $\alg$ by using the fact
that they are linear derivations and employing the Leibnitz rule.
The convention that the operators $\hat\partial_i$ generate a
central extension of the two-dimensional translation group (with
central element $\phi$) is simply for convenience later on and
will have no real bearing on the physical and geometrical objects
that we construct. This is because $\hat\partial$ acts on $\alg$
through infinitesimal automorphisms (i.e. commutators).

\subsection*{\sc 2.3.~Trace}

A trace on the algebra is a linear functional
$\Tr:\alg\to\complex$ which is cyclic with respect to its product.
Up to normalization there is a unique trace which is positive,
$\Tr\hat f^\dag\hat f\geq0~~\forall\hat f\in\alg$, and which is
compatible with the $*$-conjugation on $\alg$ in the sense that
\beq \Tr\hat f^\dag=\overline{\Tr\hat f} \ .
\label{Trconjcomp}\end{equation} It is defined on elements (\ref{hatfpower})
by picking out the zero mode in the power series expansion, \beq
\Tr\hat f=f_{\mbf0}^{~} \ . \label{Tracedef}\end{equation} This trace obeys
the ``integration by parts'' property \beq
\Tr\left[\hat\partial_i\,,\,\hat f\right]=0 \label{Trintparts}\end{equation}
with respect to the derivations constructed in (\ref{hatdeldef}). This
means that $\Tr$ is invariant under the infinitesimal action of the
(centrally-extended) translation group generated by $\hat\partial$.

\subsection*{\sc 2.4.~Fields}

To make contact with the geometrical notion of a torus
$\torus^2=\sphere^1\times\sphere^1$, we will view $\alg$ as a
deformation of the algebra $C(\torus^2)$ of functions on
$\torus^2\to\complex$. For simplicity we assume that the torus is
square and of unit area. Then there is a one-to-one correspondence
between elements of the algebra $\alg$ of the form
(\ref{hatfpower}) and functions $f:\torus^2\to\complex$ with the
two-dimensional Fourier series expansions \beq
f(x)=\sum_{\vm\in\zeds^2}f^{~}_\vm~\e^{2\pi\ii m_ix^i} \ ,
\label{Fourierseries}\end{equation} where $x=(x^1,x^2)$, $x^i\in[0,1]$ are
local coordinates on $\torus^2$. Throughout we shall use the
Einstein summation convention for implicitly summing over repeated
upper and lower indices. From this correspondence we see that,
roughly speaking, the expansion coefficients $f^{~}_\vm$ of
(\ref{hatfpower}) may be thought of as Fourier coefficients, and
the generators of $\alg$ as the plane wave basis $\hat
Z_i=\e^{2\pi\ii\hat x^i}$ for the Fourier expansion of functions
on the torus. The algebraic relations (\ref{NCtorusrels}) are then
obtained by promoting the coordinates of $\torus^2$ to Hermitian
operators $\hat x^i$ which obey the commutation relations $[\hat
x^1,\hat x^2]=\ii\theta/2\pi$, and hence a noncommutative space
appears.

More precisely, when $\theta=0$, starting from the commutative
algebra $C(\torus^2)$ (with the usual pointwise multiplication of
functions) it is possible to entirely reconstruct the torus as a
topological space. In fact, the celebrated Gel'fand-Naimark
theorem asserts that there is a one-to-one correspondence between
the category of commutative $C^*$-algebras and the category of Hausdorff
topological spaces. Thus commutative algebras correspond to
ordinary geometrical spaces, which always admit an equivalent
``dual'' description in terms of their algebras of functions (i.e.
once every function on a space is known, then so is the space
itself). When $\theta\neq0$, the algebra $\alg$ is noncommutative,
and the correspondence ceases to exist. The notion of ``space''
formally disappears in this case, but we may still keep the
underlying torus through the one-to-one correspondence described
above. Now, however, the noncommutativity of $\alg$ is manifested
through the property that, for any two functions $f$ and $g$ on
$\torus^2$, the function corresponding to the element of $\alg$
obtained by multiplying the two associated operators is given
through \beq \hat f\,\hat g=\widehat{f\star g} \ ,
\label{hatfstarg}\end{equation} and the commutative pointwise multiplication
of functions is replaced by the noncommutative star-product. There
are different ways to define the star-product by using the Fourier
series (\ref{Fourierseries}) and (\ref{hatfpower})~\cite{sz1}, but the one we
shall primarily refer back to is the integral kernel
representation \beq (f\star
g)(x)=\frac4{\theta^2}\,\int\!\!\!\int\dd^2y~\dd^2z~
f(x+y)\,g(x+z)~\e^{4\pi\ii y\wedge z/\theta} \ ,
\label{starintkernel}\end{equation} where we have introduced the
two-dimensional cross-product \beq y\wedge z=y^1z^2-y^2z^1
\label{wedgedef} \end{equation} and the integrations extend over $\torus^2$.
The star-product is associative and noncommutative, and reduces to
ordinary pointwise multiplication when $\theta=0$. The presence of
the phase oscillations in (\ref{starintkernel}) illustrates the
true non-local nature of the result of multiplying together two
fields with this product.

This is the picture that is usually employed in field theoretical
considerations, because in this manner one can treat fields on
noncommutative spaces as fields on ordinary spaces but multiplied
together using the non-local star-product. Furthermore, with
respect to this correspondence the linear derivations introduced
in (\ref{hatdeldef}) induce ordinary derivatives through \beq
\left[\hat\partial_i\,,\hat f\right]=\widehat{\partial_if}
\label{hatdelderiv}\end{equation} with $\partial_i\equiv\partial/\partial
x^i$, while the trace defined in (\ref{Tracedef}) corresponds to
integration over $\torus^2$, \beq \Tr\hat f=\int\dd^2x~f(x) \ ,
\label{Traceint}\end{equation} since by periodicity averages over $\torus^2$
also pick out zero modes of functions. The cyclic symmetry of the
trace manifests itself in the identity \beq \Tr\hat f\hat
g\equiv\int\dd^2x~(f\star g)(x)=\int\dd^2x~f(x)\,g(x)
\label{Tracecyclic}\end{equation} which follows from an integration by parts
(c.f.~(\ref{Trintparts})). The property (\ref{Tracecyclic}) also
ensures that free field actions are unaltered by noncommutativity,
which thereby only appears through the addition of interaction
terms. While we shall have occasion to refer back to this picture
for illustrative purposes, for the most part our analysis will be
done in the abstract setting without recourse to considerations of
fields on $\torus^2$ and star-products. This will prove to be the
most natural setting for solving noncommutative gauge theory in
two-dimensions. We shall therefore drop the hat label distinction
between elements of $\alg$ and fields for notational simplicity in
the following, the distinction always being clear from the context
in which we will be working.

\subsection*{\sc 2.5.~Projective Modules}

In order to define gauge theories on the noncommutative torus we
shall need the appropriate notion of a complex vector bundle over
$\alg$. Within the context described above, this is readily
accomplished by looking at the representations of the algebra
$\alg$, which from an algebraic point of view is where the real
interesting structures always lie. The reason for this
characterization follows from a classic result in bundle theory
known as the Serre-Swan theorem. Just as spaces are in a
one-to-one correspondence with commutative algebras (of
functions), so too are vector bundles in a one-to-one
correspondence with finitely-generated projective modules over the
algebras. For instance, the correspondence in the case of the
ordinary torus associates to each vector bundle $E\to\torus^2$ the
module of sections $C(\torus^2,E)\to C(\torus^2)$ of the bundle.
The space of sections clearly forms a representation of the
algebra of functions, because given any section $s:\torus^2\to E$
the map $f\cdot s$ is also a section for any function
$f:\torus^2\to\complex$, i.e. the algebra $C(\torus^2)$ acts on
$C(\torus^2,E)$.

With this algebraic characterization, we can define a vector
bundle in the noncommutative case as a finitely-generated,
projective module ${\cal E}\to\alg$ over the algebra. For
definiteness we always assume that $\alg$ acts on $\cal E$ from
the right (in the commutative case there is no distinction between
left and right actions). The stated restrictions on the module
mean that it is of the form ${\cal E}=\proj\,\alg^n$, where
$\alg^n=\alg\oplus\cdots\oplus\alg$ ($n$ times) is the free module
of rank $n$ over $\alg$, while
$\proj\in\mat_n(\alg)=\alg\otimes\mat_n$ (with $\mat_n$ the
algebra of $n\times n$ complex matrices) is a Hermitian projector
on the algebra with $\proj^2=\proj=\proj^\dag$. This condition
ensures that $\cal E$ can be embedded as a summand in a trivial
module, as then $\alg^n={\cal E}\oplus(\id-\proj)\alg^n$ with
$(\id-\proj)\alg^n$ projective. This is the noncommutative version
of the requirement spelled out in Swan's theorem, which asserts
that any complex vector bundle $E\to\torus^2$ can be embedded as a
Whitney summand in a trivial bundle $\torus^2\times\complex^n$
over the torus.

The connected components of the infinite-dimensional Grassmannian
manifold of Hermitian projectors on $\alg$ is parametrized by its
K-theory \cite{pv}
\beq
{\rm K}_0(\alg)=\zed\oplus\zed \ , \label{K0alg}
\end{equation}
defined as the group of equivalence classes of projectors modulo
stable isomorphism. The canonical trace defined in section~2.3
yields an isomorphism of ordered groups ${\rm
K}_0(\alg)\to\zed+\zed\,\theta\subset\real$ through
$\Tr\otimes\tr_n^{~}\,\proj=p-q\,\theta$, with $\tr_n^{~}$ the
usual $n\times n$ matrix trace and $\proj=\proj_{p,q}$ the
projector corresponding to the K-theory class labelled by
$(p,q)\in\zed^2$. Since $\proj$ acts as the identity on the module
${\cal E}=\proj\,\alg^n$, by positivity of the trace its dimension
obeys
\beq
\dim\module=\Tr\otimes\tr_n^{~}\,\proj=\Tr\otimes\tr_n^{~}\,\proj^\dag
\proj\geq0 \ . \label{dimpos}
\end{equation}
It follows that the stable
projective modules are classified by the positive cone of the
ordered K-theory group (\ref{K0alg}). As the module $\module$ will
be taken to be a separable Hilbert space, its dimension
(\ref{dimpos}) must be suitably defined in a regulated fashion. It
is well-known how to do this in functional analysis, and using
this it follows that to every pair of integers $(p,q)$ we can
associate a Heisenberg module $\module=\module_{p,q}$ of positive
Murray-von~Neumann dimension~\cite{Rieffel83}
\beq \dim\module_{p,q}=p-q\,\theta>0
\ . \label{dimpqpos}\end{equation}

All finitely-generated projective modules over the algebra $\alg$
are either free modules $\alg^N$, Heisenberg modules, or
combinations thereof \cite{rieffelhigher}.
The explicit representation of $\alg$
furnished by the Heisenberg modules will be described later on. To
understand geometrically the abstraction just presented, one
should simply repeat the above analysis in the more familiar case
$\theta=0$. Then (\ref{K0alg}) coincides with the topological
K-theory of the ordinary torus $\torus^2$ (K-theory groups are
stable under deformations of algebras), whose positive cone
classifies complex vector bundles over $\torus^2$ of rank $p$ and
magnetic charge $q\in\zed$. While the interpretation of $p$ as
rank generally ceases to hold when $\theta\neq0$, the integer $q$ still
corresponds to the Chern character of the ``gauge bundle'' over
$\alg$ \cite{connestransl}.

\subsection*{\sc 2.6.~Gauge Fields}

To introduce gauge fields in analogy with ordinary geometry, we
need to specify the appropriate notion of connection on a module
$\module\to\alg$. Given the way we defined derivatives in
section~2.2, we define a connection $\nabla$ to be a
``representation'' of the linear derivations $\partial$
corresponding to the module $\module$, or more precisely as a
vector space homomorphism
$\nabla:\module\to\module\otimes\heisen^*$, where here and in the
following the direct product is implicitly understood to be
$\alg$-linear. Its components satisfy the commutation relations
\beq [\nabla_i,Z_j]=2\pi\ii\,\delta_{ij}\,Z_j \ ,
\label{nablacommrels}\end{equation} which when represented on $\module$
corresponds to the usual Leibnitz rule. From (\ref{hatdeldef}) and
(\ref{nablacommrels}) it follows that $\nabla-\partial$ commutes
with all elements of $\alg$, and as a consequence we may write the
connection as a covariant derivative \beq \nabla=\partial+A \ .
\label{nabladelA}\end{equation} More precisely, the derivations in
(\ref{nabladelA}) should be written as $\proj\,\partial\,\proj$
with $\partial$ extended to $\alg^n$ in the obvious way. The gauge
fields $A_i\in{\rm
  End}(\module)=\module^*\otimes\module$ are elements of the algebra
of $\alg$-linear endomorphisms of the module $\module$ (again with
the $\alg$-linearity not indicated explicitly in the notation for
brevity), which is the commutant of the algebra $\alg$ in
$\module$ consisting of those elements of the algebra
representation which commute with all elements of~$\alg$.

As in the commutative case, we will work only with compatible
connections. The compatibility condition may be stated by
introducing an $\alg$-valued inner product on $\module$ which is
compatible with its $\alg$-module structure~\cite{connesbook}. We will
not write this condition explicitly, as it will not be required
in the following. All we shall need to know is that compatibility
implies that the curvature of the connection $\nabla$ commutes
with all elements of $\alg$, and therefore that
$[\nabla_1,\nabla_2]\in{\rm
  End}(\module)$. From a physical standpoint, it is this restriction
that allows the construction of gauge field theory actions from
the algebraic objects above. From this curvature one defines the
noncommutative field strength $F_A$ of $A$ through \beq
[\nabla_1,\nabla_2]=[\partial_1,A_2]-[\partial_2,A_1]+[A_1,A_2]+\phi
\equiv F_A+\phi \ . \label{fieldstrengthdef}\end{equation} The space of
compatible connections on an $\alg$-module $\module$ will be
denoted ${\cal C}={\cal C}(\module)$.

\subsection*{\sc 2.7.~Differential Forms}

Starting from the commutant of the algebra $\alg$ in a given
module $\module$, and the exterior products of the dual Heisenberg
Lie algebra, we can define a graded differential algebra \beq
\Omega(\module)=\bigoplus_{n\geq0}\Omega^n(\module) \ , ~~
\Omega^n(\module)={\rm
End}(\module)\otimes{\bigwedge}^n\,\heisen^* \ .
\label{diffalg}\end{equation} Elements of (\ref{diffalg}) are (left)
translationally-invariant differential forms on the Lie group
$\exp\heisen$ with coefficients in ${\rm End}(\module)$. For
example, the curvature form $[\nabla,\nabla]\in\Omega^2(\module)$.
The space of compatible connections $\cal C$ is an affine space
over the vector space ${\rm End}(\module)\otimes\heisen^*$ of
linear maps $\heisen\to{\rm End}(\module)$, whose tangent space
may thereby be identified with \beq \Omega^1(\module)={\rm
End}(\module)\otimes{\bigwedge}^1\,\heisen^* \ .
\label{tancompconn}\end{equation} Given this notion of cotangent vector, we
may then define functional differentiation of any functional
$f[A]$ at a point $\nabla=\partial+A\in{\cal C}$ through \beq
\frac\delta{\delta A}f[a]=\left.\frac\partial{\partial
t}f[A+t\,a]\right|_{t=0} \ , ~~ a\in\Omega^1(\module) \ .
\label{functdiff}\end{equation}

\subsection*{\sc 2.8.~Constant Curvature Connections}

As in the case of vector bundles over the torus $\torus^2$, in the
noncommutative case every Heisenberg module
$\module=\module_{p,q}$ admits a constant curvature connection
$\nabla^{\rm c}\in{\cal C}$, for which \beq \left[\nabla_1^{\rm
c}\,,\,\nabla_2^{\rm c}\right]=\ii f \label{constcurvf}\end{equation} with
$f$ a real constant. At the algebraic level, by constant we mean
that the curvature (\ref{constcurvf}) is proportional to identity
operator on $\module\to\module$, while at the level of fields we
mean that it is independent of the coordinates of $\torus^2$. The
constant curvature connections are the natural extensions of the
derivations defined by (\ref{hatdeldef}). They allow us to
explicitly construct the corresponding representation Hilbert
spaces.

For $q=0$ we define $\module$ to be the ($L^2$-completion of the)
free module of rank $p$ over $C(\torus^2)$,
\beq
\module_{p,0}=L^2\left(\torus^2\right)\otimes\complex^p \ ,
\label{Ep0}
\end{equation}
with the action of $\alg$ defined as
multiplication with the star-product. For $q\neq0$ we define
\beq
\module_{p,q}=L^2(\real)\otimes\complex^q \ , ~~ q\neq0 \ .
\label{Epq}
\end{equation}
The Hilbert space $L^2(\real)$ is the
Schr\"odinger representation of the Heisenberg commutation
relations (\ref{constcurvf}), which by the Stone-von~Neumann
theorem is the unique irreducible representation of the Lie
algebra $\heisen_f$. The finite dimensional vector space
$\complex^q$ carries the $q\times q$ representation of the
Weyl-'t~Hooft algebra \beq \Gamma_1\,\Gamma_2=\e^{2\pi\ii
p/q}~\Gamma_2\,\Gamma_1 \label{WeyltHooft}\end{equation} defined by \beq
\Gamma_1=\left(W_q^\dag\right)^{2p} \ , ~~
\Gamma_2=\Bigl(V_q\Bigr)^p \ . \label{Gamma12sol}\end{equation} The
traceless $SU(q)$ shift and clock matrices \bea
V_q&=&\begin{pmatrix}0&1& & &0\\ &0&1& & \\ & &\ddots&\ddots& \\
 & & &\ddots&1\\1& & & &0\end{pmatrix} \ , \non
W_q&=&\begin{pmatrix}1& & & &0\\ &\e^{2\pi\ii/q}& & & \\ &
&\e^{4\pi\ii/q}&
 & \\ & & &\ddots& \\0& & & &\e^{2\pi\ii(q-1)/q}\end{pmatrix}
\label{clockshiftdef}\eea obey the commutation relation \beq
V_q\,W_q=\e^{2\pi\ii/q}~W_q\,V_q \ . \label{clockshiftcomm}\end{equation}

We may then represent the generators of the noncommutative torus
on (\ref{Epq}) as \beq Z_i=\e^{2\pi\,\nabla_i^{\rm
c}/f}\otimes\Gamma_i \ . \label{NCgensrep}\end{equation} By working out the
commutation relation of the operators (\ref{NCgensrep}) using
(\ref{constcurvf}) and (\ref{WeyltHooft}), and comparing with
(\ref{NCtorusrels}), we arrive at an expression for $\theta$ as a
function of $f$, $p$ and $q$. This determines the constant
curvature (\ref{constcurvf}) of the given Heisenberg module
$\module_{p,q}$ in terms of its topological numbers and the
noncommutativity parameter as \beq f=\frac{2\pi\,q}{p-q\,\theta} \
. \label{fpqtheta}\end{equation} As we did in section~2.4, it is possible to
alternatively construct the Heisenberg modules in terms of fields
on the ordinary torus $\torus^2$~\cite{ks,sz1}. This gives a description of
$\module_{p,q}$ as a deformation of the $C(\torus^2)$-module of
sections of a complex vector bundle $E_{p,q}\to\torus^2$ of rank
$p$, topological charge $q$, and constant curvature $2\pi\,q/p$.
However, in what follows we shall find it more convenient to work
in the Hilbert space picture above.

Up to $SU(q)$ equivalence, the Weyl-'t~Hooft algebra
(\ref{WeyltHooft}) is known to possess a unique irreducible
representation of dimension $q/{\rm gcd}(p,q)$ \cite{vanBaal,lebedev}.
It follows that
the Heisenberg module (\ref{Epq}) decomposes into irreducible
$\alg$-modules as
\bea
\module_{p,q}&=&\underbrace{\module_{p,q}'\oplus\cdots\oplus
\module_{p,q}'\,} \ , ~~~~ \module_{p,q}'~=~L^2(\real)\otimes
\complex^{q/{\rm gcd}(p,q)} \ . \label{Epqirred}\\&&~~~~~~~~~
{\scriptstyle N}\nn
\eea
This can be used to define the rank of
the module $\module_{p,q}$ as
\beq
N={\rm gcd}(p,q) \ .
\label{rankgcd}
\end{equation}
By an explicit construction \cite{cr} it is possible to
show that the commutant of the algebra $\alg=\alg_\theta$ in the
irreducible Heisenberg module is another noncommutative torus
${\rm
  End}(\module_{p,q}')\cong\alg_{\theta'}$, where
\beq \theta'=\frac{n-s\,\theta}{p-q\,\theta}\,N
\label{dualthetaN}\end{equation} is a {\it dual} noncommutativity parameter,
with the integers $n,s\in\zed$ satisfying the Diophantine property
\beq p\,s-q\,n=N \ . \label{Diophantine}\end{equation} We shall describe
this duality in more detail in section~5. It follows that the
endomorphism algebra of (\ref{Epqirred}) is given by
\beq
{\rm
End}\left(\module_{p,q}\right)=\mat_N(\alg_{\theta'}) \ .
\label{Endisoprime}
\end{equation}
This fact will enable us to identify gauge
fields of the $\alg_\theta$-module as ordinary gauge fields on
$\torus^2$ multiplied together with a star-product defined by the
dual noncommutativity parameter (\ref{dualthetaN}). Furthermore,
the canonical trace $\Tr$ on $\alg$ thereby induces a trace
$\Tr\otimes\tr^{~}_N$ on (\ref{Endisoprime}). With a slight abuse
of notation, we shall also denote this trace by $\Tr$ as it will
be contextually clear which one we mean.

\newsection*{\sc 3.~Gauge Theory on the Noncommutative Torus}

Having dispensed with the mathematical formalities, we will now
define classical Yang-Mills theory on the noncommutative torus. We
will argue in this section, at the classical level, that this
noncommutative field theory is an exactly solvable model. We fix a
Heisenberg module $\module=\module_{p,q}$ over the algebra $\alg$
and consider the affine space $\cal C$ of all compatible
connections on $\module$. In string theory, this field theory
describes the low-energy dynamics of a bound state configuration
of $p$ coincident D2-branes which carry $q$ units of D0-brane
vortex charge, in a background $B$-field \cite{sw}.

\subsection*{\sc 3.1.~Definition}

For any $\nabla=\partial+A\in{\cal C}$, we define the
noncommutative Yang-Mills action \beq
S[A]=S[\nabla]=\frac1{2g^2}\,\Tr[\nabla_1,\nabla_2]^2 \ ,
\label{YMactiondef}\end{equation} where $g$ is the (dimensionless)
Yang-Mills coupling constant. We will regard (\ref{YMactiondef})
as a functional $S:{\cal C}\to\real_+$. By using the
operator-field correspondence described in section~2.4, the
formula (\ref{fieldstrengthdef}) and the discussion at the end of
section~2.8, we may write (\ref{YMactiondef}) in a more
conventional field theoretic form as \beq
S[A]=\frac1{2g^2}\,\int\dd^2x~\tr^{~}_N\Bigl(F_A(x)+\phi\Bigr)^2 \
, \label{YMactiongauge}\end{equation} where \beq
F_A=\partial_1A_2-\partial_2A_1+A_1\star A_2-A_2\star A_1
\label{NCfieldstrength}\end{equation} is the noncommutative field strength
of an anti-Hermitian $U(N)$ gauge field $A_i(x)$ on the torus
$\torus^2$. The star-product in (\ref{NCfieldstrength}) is the
tensor product of the associative star-product corresponding to
the dual noncommutativity parameter (\ref{dualthetaN}) with
ordinary matrix multiplication. This extended star-product is
still associative.

Due to the isomorphism (\ref{Endisoprime}), in the case of gauge
theory on a Heisenberg module the action may be represented as in
(\ref{YMactiongauge}) in terms of star-products of fields on
$\torus^2$. This is {\it not} possible on a generic projective
module over the noncommutative torus, in which case one can
generally only use the more abstract definition
(\ref{YMactiondef}) of the field theory. Note that the action
(\ref{YMactiongauge}) is quadratic in the field strength and so
the non-local interactions due to the star-product appear
explicitly only in (\ref{NCfieldstrength}). This is one of the
properties that will enable an exact solution of the field theory
later on. In (\ref{YMactiongauge}) we also see that the central
extension in (\ref{hatdeldef}) has the physical interpretation of
a constant background magnetic flux in the noncommutative field
theory, and thereby plays no role in the local dynamics.

\subsection*{\sc 3.2.~Gauge Symmetry}

The action (\ref{YMactiondef}) is invariant under any covariant
transformation of the form \beq
\nabla~\longmapsto~U\,\nabla\,U^\dag \ , \label{covtransf}\end{equation}
where $U\in{\rm End}(\module)$ is a unitary operator on $\module$,
\beq U^\dag\,U=U\,U^\dag=\id \ . \label{unitarygauge}\end{equation} The set
of transformations (\ref{covtransf}) form the gauge group ${\cal
G}={\cal G}({\cal E})$ of the noncommutative gauge theory.
Infinitesimally, it follows from (\ref{covtransf}) that the gauge
group acts on gauge potentials through \beq
A~\longmapsto~A+\delta_\lambda A \ , ~~ \delta_\lambda A=[\nabla,
\lambda] \ , \label{inftransf}\end{equation} where $\lambda\in{\rm
End}(\module)$ is anti-Hermitian. The commutator of two such gauge
transformations gives \beq
\left[\delta_\lambda\,,\,\delta_{\lambda'}\right]A=\delta_{[\lambda\,,\,
\lambda']}A \ , \label{infcomm}\end{equation} and as a consequence the gauge
group $\cal G$ acts on $\cal C$. Furthermore, the curvature
transforms covariantly under the action of $\cal G$ as \beq
\delta_\lambda F_A=\left[\lambda\,,\,F_A\right] \ .
\label{Fcovtransf}\end{equation} In fact, any element of the differential
algebra (\ref{diffalg}) obeys this homogeneous transformation law,
\beq \delta_\lambda a=[\lambda,a] ~~~~\forall
a\in\Omega^n(\module),n\geq0 \ , \label{deltalambdadiff}\end{equation}
because of the lifting of the space of connections $\cal C$ to the
vector spaces of differential forms through the endomorphism
algebra ${\rm End}(\module)$.

Under the operator-field correspondence, we can write
(\ref{inftransf}) in terms of functions on the torus $\torus^2$ as
\beq \delta_\lambda A_i=\partial_i\lambda+\lambda\star A_i-A_i
\star\lambda \ , \label{inftransfT2}\end{equation} where $\lambda(x)$ is a
smooth, anti-Hermitian $N\times N$ matrix-valued field on
$\torus^2$. From (\ref{inftransfT2}) it follows that
noncommutative gauge transformations mix colour and spacetime
degrees of freedom together in a non-trivial way. This is a simple
way to see the fact that there is no notion of a structure group
in noncommutative gauge theory, only the gauge group $\cal G$. In
the commutative case, the gauge group would be the direct product
of the finite-dimensional structure group with the group of
$\sphere^1$-valued functions, and the separation of spacetime and
internal gauge symmetries would be clear. This is not so in the
noncommutative case, and we are forced to deal with spacetime and
colour symmetries on the same level.

This fact suggests that certain transformations
(\ref{inftransfT2}) could induce spacetime transformations which
manifest themselves as internal gauge symmetries of the
noncommutative field theory. In fact, noncommutative gauge
transformations have a deep geometrical interpretation \cite{lsz}. The gauge
parameters $\lambda$ live in a Lie algebra that generates the
gauge group. A basis for the spacetime part of this Lie algebra is
provided by the generators $Z_i'$ of the algebra $\alg_{\theta'}$,
with the $\lambda$'s given as appropriate anti-Hermitian
combinations. Motivated by the Fourier series expansion
(\ref{hatfpower}), we introduce the basis of operators $T_\vn$,
$\vn=(n_1,n_2)\in\zed^2$ defined by \beq
T_\vn=\frac1{\theta'}~\e^{\ii
  n_1n_2\theta'/2}~Z_1'^{\,n_1}\,Z_2'^{\,n_2} \ ,
\label{Tvndef}\end{equation} which satisfy the conjugation relation
$T_\vn^\dag=T_{-\vn}$. The Fourier coefficients of a gauge
parameter $\lambda$ thereby obey
$\lambda_{-\vn}=-\,\overline{\lambda_\vn}$. By using
(\ref{NCtorusrels}) it is straightforward to compute that the
elements (\ref{Tvndef}) close the infinite-dimensional Lie algebra
\beq
\left[T_\vn\,,\,T_\vm\right]=\frac{2\ii}{\theta'}\,\sin\left(\theta'\,
\vn\wedge\vm\right)~T_{\vn+\vm} \ , \label{Tvnalg}\end{equation} where the
two-dimensional cross product of vectors in $\zed^2$ is defined as
in (\ref{wedgedef}). The commutation relations (\ref{Tvnalg}) are
familiar from condensed matter physics, where they correspond to
the algebra of magnetic translation operators for electrons moving
in two-dimensions under the influence of a constant
perpendicularly applied magnetic field, projected onto the lowest
Landau level~\cite{Jackiw}. This coincidence is not an accident, because
noncommutative gauge theory arises most naturally as the
low-energy effective field theory of open string modes on a
D-brane, whose worldvolume is subjected to a constant background
magnetic field $B\sim\theta'^{\,-1}$~\cite{sw}. The geometrical significance
of this algebra may be seen by working close to the commutative
limit $\theta'\to0$ of the noncommutative gauge theory. With the
gauge generators (\ref{Tvndef}) denoted $T_\vn^0$ in this limit,
(\ref{Tvnalg}) reduces to \beq
\left[T_\vn^0\,,\,T_\vm^0\right]=2\ii\vn\wedge\vm~T^0_{\vn+\vm} \
. \label{Tvnalg0}\end{equation} This Lie algebra is recognized as the
classical $W_\infty$ algebra of area-preserving diffeomorphisms of
$\torus^2$.

We conclude that the Lie algebra (\ref{Tvnalg}) of noncommutative
gauge transformations is a trigonometric deformation of the
algebra $w_\infty(\torus^2)$ of area-preserving diffeomorphisms of
the torus in two-dimensions~\cite{lsz,sheikhren}. This leads to the
first indication that noncommutative Yang-Mills theory in two
spacetime dimensions is an exactly solvable model. While the area-preserving
transformations do not exhaust all two-dimensional
diffeomorphisms, they kill enough of the local degrees of freedom
to eliminate all propagating modes, and there are no gluons in
this gauge theory. In other words, noncommutative gauge theory in
two dimensions is ``almost'' a topological field theory, because
its gauge symmetry ``almost'' includes general covariance. While a
similar argumentation also follows through in the case of ordinary
Yang-Mills theory on $\torus^2$~\cite{cmr}, the situation here is more
drastic. While in the commutative case the area-preserving
symmetry is manifested through the geometrical invariance
properties of the integrated two-form field strength and as such
defines an outer automorphism of the algebra of functions
$C(\torus^2)$, here the symmetry manifests itself as an inner
automorphism of the algebra $\alg_{\theta'}$.

This argument also illustrates the crucial difference between two
dimensions and higher spacetime dimensions. While the analogous
argument in $d$ even dimensions would lead one to the conclusion
that the gauge symmetries contain symplectic diffeomorphisms~\cite{lsz}, for
$d>2$ the algebra of symplectomorphisms is much smaller than that
of volume-preserving diffeomorphisms, and only for $d=2$ do the
canonical transformations coincide with area-perserving
diffeomorphisms. These same conclusions can also be reached by
noting that locally one can fix an axial gauge to completely
eliminate the star-product interaction terms from the action
(\ref{YMactiongauge}), and naively one would conclude that theory
is just a trivial, free field Gaussian model. Insofar as the
computation of vacuum amplitudes is concerned, this would
certainly be true on the plane $\real^2$, but on the torus such a
gauge choice is incompatible with topologically non-trivial gauge
transformations which wind around its cycles. Because of this fact
though, the characteristics of the theory will depend only on
topological quantities associated with global degrees of freedom
of the gauge theory. We will see this explicitly throughout these
notes.

We can give the noncommutative gauge group another interpretation
which is somewhat more algebraic in nature. Suppose that we
approximate the noncommutativity parameter by a sequence of rational
numbers $m/n$, with $m,n\to\infty$ relatively prime positive
integers such that $\theta'=\lim_{m,n}m/n$ is finite. Upon
substituting $\theta'=m/n$ into (\ref{Tvnalg}) we observe that
this Lie algebra has a finite-dimensional, $n\times n$ unitary
representation in terms of clock and shift matrices
(c.f.~(\ref{WeyltHooft})--(\ref{clockshiftcomm})). These operators
span the Lie algebra of traceless Hermitian $n\times n$ matrices
and form the Fairlie-Fletcher-Zachos trigonometric basis
for $su(n)$~\cite{FFZ}. The limit $n\to\infty$ thereby identifies
the gauge group of noncommutative Yang-Mills theory as a certain
completion of the infinite unitary group
$U(\infty)$~\cite{lsz}. Physically, this infinite-dimensional symmetry
arises from the infinitely-many image D-branes associated with open
string modes terminating on a toroidal worldvolume in a
$B$-field~\cite{Cornalba0}--\cite{HKL}. The rigorous definition of
the $n\to\infty$ limit may be found in~\cite{lls1}. We shall see
how this result can be derived in another way in the second part
of these lecture notes. We remark that this conclusion is only
valid locally on $\torus^2$, because the torus $W_\infty$ algebra
requires a central extension in order to make contact with its
group theoretical description, and the torus unitary group has
only a semi-infinite Dynkin diagram~\cite{lsz}.

In the present case, the completed unitary symmetry group
$\overline{U(\infty)}$ consists of those gauge transformations
(\ref{covtransf}) whereby $U=\id+K$, with $K$ an element of the
algebra of compact endomorphisms of the module $\module$ which may
be defined as the operator norm closure of the algebra of finite-rank
endomorphisms~\cite{inprep},\cite{nair}--\cite{schwarzEucl}. The
latter algebra forms a self-adjoint
two-sided ideal in ${\rm End}(\module)$ which, since $\module$ is
separable, is isomorphic to the infinite matrix algebra
$\mat_\infty$. What is particularly interesting in this respect is
that, by Palais' theorem \cite{palais}, the completed group
$\overline{U(\infty)}$ has the same homotopy type as the finite
rank group $U(\infty)$, all of whose homotopy groups are known. We
can thereby write down explicitly the homotopy groups of the
noncommutative gauge group as \beq
\pi_n\Bigl(\,\overline{U(\infty)}\,\Bigr)=\pi_n\Bigl(U(\infty)
\Bigr)=\left\{~\begin{matrix}\zed&,&n~~{\rm odd}\\0&,&n~~{\rm
even}
\end{matrix}\right. \ .
\label{NChomotopy}\end{equation} This enables one to classify objects in the
noncommutative gauge theory which have a direct topological
characterization, such as anomalies and solitonic
configurations~\cite{harvey}.

\subsection*{\sc 3.3.~Integrability}

The fact that noncommutative gauge theory in two dimensions is an
exactly solvable model is most likely to be connected with the fact
that it is an integrable system. We will now indicate how this may
indeed be the case by showing that Yang-Mills theory on the
projective module $\module$ naturally defines an
infinite-dimensional Hamiltonian system~\cite{inprep}. On the affine space
${\cal C}(\module)$, with tangent space identified as
(\ref{tancompconn}), we can define a natural symplectic structure
by the two-form \beq \omega[a,b]=\Tr a\wedge b \ , ~~
a,b\in\Omega^1(\module) \ . \label{omegadef}\end{equation} This form is
clearly non-degenerate. It is also independent of the point
$\nabla=\partial+A\in{\cal C}$ at which it is evaluated, and hence
it is closed: $\delta\omega/\delta A=0$. The important feature of
the symplectic form (\ref{omegadef}) is that it is gauge
invariant. Using the transformation rule (\ref{deltalambdadiff})
and cyclicity of the trace, we easily find \beq
\omega\left[\delta_\lambda a\,,\,\delta_\lambda
b\right]=\omega[a,b] \ . \label{omegagaugeinv}\end{equation}

It follows that the gauge group $\cal G$ acts symplectically on
the affine space $\cal C$. Since $\cal C$ is contractible, there
exists a corresponding moment map $\mu:{\cal C}\to{\rm
End}(\module)^*$ which naturally generates a system of Hamiltonian
functionals $H_\lambda:{\cal
  C}\to\real$, $\lambda\in{\rm End}(\module)$ through
\beq
H_\lambda[A]=\Tr\mu[A]\lambda \ .
\label{Hammomentdef}
\end{equation}
As we show below, the moment map in this case is nothing but the
noncommutative field strength
\beq
\mu[A]=F_A+\phi \ ,
\label{mufieldstrength}
\end{equation}
and so the Yang-Mills action
(\ref{YMactiondef}) is the {\it square} of the moment map. The
action of this Hamiltonian system is thereby given by the square
of its Hamiltonian, rather than by just the Hamiltonian itself.
Since $\pi_2(U(\infty))=0$, the map $\lambda\mapsto H_\lambda$ is
a Lie algebra homomorphism from ${\rm
  End}(\module)$ into the Poisson algebra on $\cal C$ induced by the
symplectic structure (\ref{omegadef}). Whether or not this is
enough to construct the infinitely many conserved charges required
for complete integrability of the noncommutative Yang-Mills system
is not presently understood.

To prove (\ref{mufieldstrength}), we note that near a point
$\nabla=\partial+A\in{\cal C}$ we have \beq
F_{A+t\,a}=F_A+t\,\left[\nabla\,\stackrel{\wedge}{,}\,a
\right]+O\left(t^2\right) \label{Fnearnabla}\end{equation} where
$a\in\Omega^1(\module)$ and $t\to0$. By using the definition
(\ref{functdiff}), we may then compute the functional derivative
\bea \frac\delta{\delta A}H_\lambda[a]&=&\frac\delta{\delta
A}\Tr\left(F_A+
\phi\right)\lambda\non&=&\Tr\left[\nabla\,\stackrel{\wedge}{,}\,a
\right]\lambda\non&=&-\Tr[\nabla,\lambda]\wedge a \ ,
\label{functdiffHam}\eea where we have used the Leibnitz rule and
the integration by parts property (\ref{Trintparts}) for the
connection $\nabla$. From (\ref{inftransf}) and (\ref{omegadef})
it therefore follows that \beq \frac\delta{\delta
A}H_\lambda[a]=-\omega\left[\delta_\lambda A\,,\,a \right] \ ,
\label{Hamflows}\end{equation} which are just the Hamilton equations of
motion in the present case. The existence of these flows is
equivalent to the $\cal G$-invariance (\ref{omegagaugeinv}) of the
symplectic structure. Note that the background flux $\phi$ plays
no role in this derivation and could in principle be simply
dropped from the moment map (\ref{mufieldstrength}), thereby
indicating once again its irrelevance with respect to dynamics.

\newsection*{\sc 4.~Classical Solutions}

In this section we will come to the crux of our presentation, the
exact solution of the classical noncommutative gauge theory. The
technique we will employ is essentially an algebraic version of
the Atiyah-Bott bundle splitting method \cite{AB} for obtaining the
solutions of the Yang-Mills equations on ordinary Riemann
surfaces. It relies heavily on the properties of Heisenberg
modules over the noncommutative torus that we discussed in
section~2. The exact classical solutions of noncommutative gauge theory
on the torus were first constructed in~\cite{cr,rieffel}.

\subsection*{\sc 4.1.~Equations of Motion}

The classical field theory is defined by extremizing the action
functional (\ref{YMactiondef}). We therefore seek the stationary
points of noncommuative Yang-Mills theory which are determined by
the equation \beq \frac\delta{\delta A}S[A]=0 \ .
\label{deltaS0}\end{equation} By using (\ref{functdiff}) this leads to the
noncommutative Yang-Mills equations of motion \beq
\Bigl[\nabla\,,\,\left[\nabla_1\,,\,\nabla_2\right]\Bigr]=0 \ .
\label{NCYMeoms}\end{equation} Under the operator-field correspondence,
these equations may be cast into a more conventional form as \beq
\partial_iF_A+A_i\star F_A-F_A\star A_i=0
\label{NCYMeomsspace}\end{equation} for $i=1,2$. We seek to characterize all
such critical points of noncommutative gauge theory. This we will
do within a particular homotopy class of the Grassmannian manifold
of projectors of the algebra $\alg$, characterized by a Heisenberg
module $\module=\module_{p,q}$ of topological numbers
$(p,q)\in\zed^2$ and dimension (\ref{dimpqpos}). As we will see,
in this case there is a nice topological classification of the
classical solutions on the space $\cal C$ of compatible
connections of the Heisenberg module. In addition, by combining
these results with those of section~3.3, it will follow
immediately that the Yang-Mills action (\ref{YMactiondef}) is a
gauge-equivariant Morse functional on $\cal C$.

\subsection*{\sc 4.2.~BPS States}

Recall from section~2.8 that a Heisenberg module
$\module=\module_{p,q}$ is completely characterized by a
connection $\nabla^{\rm c}\in{\cal C}$ of constant curvature
(\ref{constcurvf},\ref{fpqtheta}). Such connections trivially
satisfy the equations of motion (\ref{NCYMeoms}), because their
curvatures are proportional to the identity operator on $\module$
and are thereby central. What is striking about the constant
curvature solutions is that they provide the absolute
minimum value of the noncommutative Yang-Mills action
(\ref{YMactiondef}) on the module $\module$~\cite{cr,amns},
\beq
S\left[\nabla^{\rm c}\right]=\inf_{\nabla\in{\cal C}}\,S[\nabla] \ .
\label{Sinfemum}\eeq
They thereby constitute the stable classical vacuum states of the gauge
theory.
In an appropriate supersymmetric extension of the field theory, constant
curvature connections correspond to $\frac12$ BPS configurations
\cite{cmrMatrix,gme}.

To prove this fact, we use the expansion (\ref{Fnearnabla}) of the
field strength to expand the action (\ref{YMactiondef}) about a
constant curvature connection to get \beq S\left[\nabla^{\rm
c}+t\,a\right]=S\left[\nabla^{\rm c}\right]+
\frac{t^2}{2g^2}\,\Tr\left[\nabla^{\rm c}\,\stackrel{\wedge}{,}
\,a\right]^2+O\left(t^4\right) \label{Sconstexp}\end{equation} for $t\to0$,
where we have used integration by parts (\ref{Trintparts}) and the
equations of motion (\ref{NCYMeoms}) to eliminate the term of
order $t$ in (\ref{Sconstexp}). The order $t^2$ term is positive,
and so we conclude that \beq S\left[\nabla^{\rm c}+a\right]\geq
S\left[\nabla^{\rm c}\right] ~~~~\forall a\in\Omega^1(\module) \ .
\label{localminarg}\end{equation} To show that $\nabla^{\rm
c}=\partial+A^{\rm c}$ is a {\it global} minimum, we can exploit
the freedom of choice of the background flux $\phi$, which we
establish more rigorously in the next section, to identify it with
the constant curvature of $\module_{p,q}$ by choosing \beq
\phi=-F_{A^{\rm c}}=-\frac{2\pi\,q}{p-q\,\theta} \ .
\label{phichoice}\end{equation} Then $S[\nabla^{\rm c}]=0$, and since
(\ref{YMactiondef}) is a positive functional, the connection
$\nabla^{\rm c}$ corresponds to the ground state, as claimed.
Henceforth we shall always assume that the natural boundary
condition (\ref{phichoice}) has been chosen.

\subsection*{\sc 4.3.~Unstable Vacua}

Constant curvature connections can also be used to construct {\it
all} solutions to the classical equations of motion of
noncommutative gauge theory. Most of these stationary points will
be local maxima or saddle-points, and hence will be unstable.
Nevertheless, they are still perfectly good field configurations
and play an important role in the quantum dynamics of the gauge
theory, as we will see in the second part of these lecture notes.
Their general structure can be deduced as follows.

For any connection $\nabla=\partial+A\in{\cal C}(\module)$, the
adjoint action of the field strength $F_A$ on the graded
differential algebra (\ref{diffalg}) is generated through the
self-adjoint linear operator
$\Xi_\nabla:\Omega(\module)\to\Omega(\module)$ defined by \beq
\Xi_\nabla(\alpha)=\left[F_A\,,\,\alpha\right]
\label{Xinabladef}\end{equation} for $\alpha\in\Omega(\module)$. Let us
consider this operator for $\nabla$ near a critical point, i.e. a
solution of the equations of motion. We denote such a generic
critical point by $\nabla^{\rm
  cl}=\partial+A^{\rm cl}$. Then the Yang-Mills
equations (\ref{NCYMeoms}) can be written as \beq \left[F_{A^{\rm
cl}}\,,\,\nabla_i^{\rm cl}\right]=0 \ , ~~ i=1,2 \ .
\label{NCYMrewrite}\end{equation} This implies that the field strength
definition $F_{A^{\rm cl}}=[\nabla_1^{\rm cl},\nabla_2^{\rm cl}]$
for an on-shell gauge field corresponds (up to an irrelevant shift
by the constant curvature $\phi$) to a Heisenberg Lie algebra in
${\rm End}(\module)$, with generators $\nabla_1^{\rm cl}$,
$\nabla_2^{\rm cl}$ and $F_{A^{\rm cl}}$, and with central element
the field strength. From this fact it follows that the eigenvalues
$c_k\in\alg$ of the operator (\ref{Xinabladef}) are central
elements near $\nabla=\nabla^{\rm cl}$, i.e. they are proportional
to the unit $\id$ of $\alg$ in the algebraic setting, or
equivalently constant functions on $\torus^2$ under the
operator-field correspondence.

In particular, this means that in the neighbourhood of any
critical point there is an eigenspace decomposition
$\Omega(\module)=\bigoplus_k\Omega_k$, where the operator
$\Xi_\nabla$ acts on each $\Omega_k$ as multiplication by a fixed
scalar $c_k\in\real$. We can interpret the eigenspaces
$\Omega_k=\Omega({\cal
  E}_{p_k,q_k})$ as the differential algebras of submodules
$\module_{p_k,q_k}\subset\module_{p,q}$. Thus near a classical
solution the module $\module_{p,q}$ may be regarded as admitting a
natural direct sum decomposition into projective submodules, \beq
\module_{p,q}={\bigoplus}_k\,\module_{p_k,q_k} \ .
\label{directsumdecomp}\end{equation} We stress that (\ref{directsumdecomp})
does {\it not} mean that the given Heisenberg module of the
noncommutative gauge theory is reducible. It simply reflects the
behaviour of connections on $\module_{p,q}$ near a stationary
point of the noncommutative Yang-Mills action. Given a collection
of connections $\nabla_{(k)}$ on $\module_{p_k,q_k}$, the action
functional (\ref{YMactiondef}) possesses the additivity property
\beq
S\left[{\bigoplus}_k\,\nabla_{(k)}\right]={\sum}_k\,S\left[\nabla_{(k)}\right]
\label{YMadditive}\end{equation} with respect to the decomposition
(\ref{directsumdecomp}).

{}From (\ref{NCYMrewrite}) it also follows that each critical
connection preserves the submodules, $\nabla_i^{\rm
  cl}:\module_{p_k,q_k}\to\module_{p_k,q_k}$, and hence we may define
a connection $\nabla_{(k)}^{\rm c}=\nabla^{\rm
  cl}|_{\module_{p_k,q_k}}$ of constant curvature $F_{A^{\rm
    cl}}|_{\module_{p_k,q_k}}$ on each $\module_{p_k,q_k}$. From the
results of the previous subsection, we know that each term
$S[\nabla_{(k)}]$ on the right-hand side of (\ref{YMadditive}) is
minimized by $\nabla_{(k)}^{\rm c}$. It thus follows from
(\ref{YMadditive}) that the noncommutative Yang-Mills action has a
critical point \beq \nabla^{\rm
cl}={\bigoplus}_k\,\nabla_{(k)}^{\rm c} \label{NCYMcritpt}\end{equation} on
the module splitting (\ref{directsumdecomp}).

We have thereby deduced that {\it every} classical solution of
noncommutative gauge theory is of the form (\ref{NCYMcritpt}) and
is characterized by a submodule decomposition
(\ref{directsumdecomp}) of the original Heisenberg module.
Generically, there could be many different types of splittings
(\ref{directsumdecomp}), but there are two important constraints
which must be met:
\begin{itemize}
\item {\it Dimension additivity:} Since (\ref{directsumdecomp}) is a
  direct sum decomposition, we have
  $\dim\module_{p,q}=\sum_k\dim\module_{p_k,q_k}$, or equivalently
  $p-q\,\theta=\sum_k(p_k-q_k\theta)$.
\item {\it K-theory charge conservation:} Demanding in addition that the Chern
  number be additive, $q=\sum_kq_k$, implies
  $(p,q)=\sum_k(p_k,q_k)$.
\end{itemize}
These conservation laws are very natural in the string theory
setting, in which the positive cone of the K-theory group
(\ref{K0alg}) represents the full lattice of D2--D0 brane
charges. Note that locally any classical solution $\nabla^{\rm cl}$
may be regarded as a particular (non-BPS) combination of BPS field
configurations.

\subsection*{\sc 4.4.~Partitions}

The conclusions just arrived at in the previous subsection show
that, for all $\theta$, any solution of the classical equations of
motion for Yang-Mills theory on the projective module
$\module_{p,q}$ is completely characterized
by a {\it partition}~\cite{rieffel,inprep}, which we define as a
collection of integers $(\vp,\vq)=\{(p_k,q_k)\}$ satisfying the constraints
\bea
p_k-q_k\theta&>&0 \ , \non{\sum}_k\,\left(p_k-q_k\theta\right)&=&
p-q\,\theta \ , \non{\sum}_k\,q_k&=&q \ .
\label{partitiondef}
\eea
In addition, to avoid overcounting, it is sometimes useful to
impose a further ordering constraint $p_k-q_k\theta\leq
p_{k+1}-q_{k+1}\theta~~\forall k$, and regard any two partitions
as the same if they coincide after rearranging their components
according to this ordering. We may then characterize the
components of a partition by integers $\nu_a>0$ which are defined as the
number of partition components that have the $a^{\rm th}$ least
dimension $p_a-q_a\theta$. The integer \beq |\vnu|={\sum}_a\,\nu_a
\label{vnudef}\end{equation} is then the total number of components in the
given partition. Note that if $\theta$ is an irrational number,
then the constraint on the magnetic charges in
(\ref{partitiondef}) is automatically ensured by the dimension
additivity. This is not the case, however, for rational-valued
noncommutativity parameters. In particular, in the commutative
case $\theta=0$ this last constraint distinguishes between
``physical'' Yang-Mills theory, which would take into account the
sum over all topological charges $q\in\zed$ (i.e. all isomorphism
classes of principal $U(p)$ bundles over $\torus^2$), and
Yang-Mills theory defined on a particular projective module. In
the noncommutative setting, only gauge theory on a fixed
projective module appears amenable to an unambiguous definition,
reflecting the loss of structure group in this case.

With this at hand, we can finally evaluate the noncommutative
Yang-Mills action on a classical solution, with corresponding
partition $(\vp,\vq)$. By using the additivity property
(\ref{YMadditive}) of the action, the formula (\ref{fpqtheta}) for
the constant curvature of a Heisenberg module, and the
Murray-von~Neumann dimension formula
$\Tr\id|_{\module_{p_k,q_k}}=p_k-q_k\theta$, we easily arrive at
\bea S(\vp,\vq)&=&S\left[{\bigoplus}_k\,\nabla^{\rm
c}_{(k)}\right] \non
&=&\frac{2\pi^2}{g^2}\,{\sum}_k\,\left(p_k-q_k\theta\right)
\left(\frac{q_k}{p_k-q_k\theta}-\frac q{p-q\,\theta}\right)^2 \ .
\label{NCYMpartition}\eea From the classical action one can in
fact determine a number of important facts about the critical
point set of noncommutative gauge theory. First of all, it is
straightforward to see that, up to gauge equivalence, the action
(\ref{NCYMpartition}) vanishes {\it only} for the trivial
partition $(\vp,\vq)=\{(p,q)\}$ corresponding to the Heisenberg module
$\module_{p,q}$ itself. This is equivalent to the statement that
the gauge theory is globally minimized by its constant curvature
connection $\nabla^{\rm c}$, as we showed in section~4.2.
Secondly, for any fixed finite action solution of the
noncommutative Yang-Mills equations of motion, it is
straightforward to show from (\ref{NCYMpartition}) that each
partition contains finitely many components~\cite{inprep}. This shows
that the number (\ref{vnudef}) is indeed finite and permits one
to pick out a minimum dimension submodule $\module_{p_1,q_1}$ as
explained above. Finally, one can show that the set of values of
the noncommutative Yang-Mills action on the critical point set
(the set of all partitions) is discrete~\cite{rieffel}, so that the
action indeed defines a bonafide Morse functional on ${\cal
C}(\module_{p,q})$.

\newsection*{\sc 5.~Morita Duality}

In noncommutative geometry an important role is played by the
notion of ``Morita equivalence'' of $C^*$-algebras. Two algebras
are Morita equivalent if they have the same representation theory.
The one-to-one correspondence between their projective modules
means in particular that they have the same K-theory. In
section~2.8 we in fact encountered a very important case of this
duality for noncommutative tori. Namely, the two algebras
$\alg_\theta$ and $\alg_{\theta'}$, with noncommutativity
parameters related through (\ref{dualthetaN}), are both
represented on the Heisenberg module $\module=\module_{p,q}$ and
commute with each other in the representation. They are therefore
Morita equivalent and in this context $\module$ is refered to as a
Morita equivalence bimodule between $\alg_\theta$ and
$\alg_{\theta'}$.

Given that noncommutative gauge theories are constructed from the
representations of the algebra, this equivalence has the potential
of supplying a novel duality relation between field theories
defined on different noncommutative spaces. In fact, this is one
of the reasons for the excitement about noncommutative Yang-Mills
theory. The Morita equivalence between two noncommutative tori
turns out to be equivalent to the fact that the underlying tori
$\torus^2$ are related by open string T-duality~\cite{gme,lls2}. The
noncommutative Yang-Mills action is invariant under the
equivalence \cite{amns}, provided we use an extended notion of Morita
equivalence known as gauge Morita equivalence. Under the extended
equivalence, not only do we obtain a one-to-one correspondence
between projective modules over different noncommutative tori
associated with different topological numbers, but we also augment
this with transformations of connections between the modules. The
conclusion is then that noncommutative Yang-Mills theory is {\it
invariant} under T-duality \cite{sw,schwarzPhi}.
This is in marked contrast to the
situation in ordinary Yang-Mills theory which is not duality
invariant. T-duality invariance is one of the remarkable features
of noncommutative field theories that follow from their origins in
string theory, yet can be defined and analysed within the
framework of quantum field theory. In this final section we will
present the Morita transformation rules and describe various
consequences of them.

The full open string T-duality group on $\torus^2$ is
$SO(2,2,\zed)$ and it acts on the K-theory ring ${\rm
K}_0(\alg)\oplus{\rm K}_1(\alg)$ in a spinor representation, in
precisely the same way in which it acts on the Ramond-Ramond
charges of D-branes. In generic spacetime dimension $d$ this group
is $SO(d,d,\zed)$. The special feature of two dimensions is that
there is an isomorphism
\beq
SO(2,2,\zed)=SL(2,\zed)\times
SL(2,\zed) \ . \label{SO22SLs}
\end{equation}
One of these $SL(2,\zed)$
factors is just the geometrical automorphism group of the torus
$\torus^2$ which acts by discrete M\"obius transformations of its
Teichm\"uller modulus $\tau$. It is the same symmetry group that
is present in the commutative case and so will not be discussed
any further. The other $SL(2,\zed)$ acts on the K\"ahler modulus
of $\torus^2$ and has no counterpart in ordinary geometry. It acts
by discrete M\"obius transformations of the noncommutativity
parameter $\theta$. In fact, a well-known theorem of
noncommutative geometry \cite{Rieffel83}
asserts that two noncommutative tori
$\alg=\alg_\theta$ and $\alg'=\alg_{\theta'}$ are Morita
equivalent if and only if \beq
\theta'=\frac{m\,\theta+n}{r\,\theta+s} \ , ~~ \begin{pmatrix}
m&n\\r&s\end{pmatrix} \in SL(2,\zed) \ . \label{thetaMorita}\end{equation}
This coincides with the transformation rule for the quantity $G+B$
under T-duality~\cite{sw}, where $G$ is the open string metric and $B$ the
antisymmetric tensor field.

The same transformation sends a projective module $\module$ of
topological numbers $(p,q)$ over the algebra $\alg$ to a module
$\module'$ of topological numbers $(p',q')$ over $\alg'$, in such
a way that the pair of K-theory charges transforms as an
$SL(2,\zed)$ doublet \beq
\begin{pmatrix}p'\\q'\end{pmatrix}=\begin{pmatrix}
m&n\\r&s\end{pmatrix}\begin{pmatrix}p\\q\end{pmatrix} \ .
\label{pqdoublet}\end{equation} From (\ref{thetaMorita}) and
(\ref{pqdoublet}) it is straightforward to show that the
Murray-von~Neumann dimensions (\ref{dimpqpos}) of the modules are
related by \beq \dim\module'=\frac{\dim\module}{|r\,\theta+s|} \ ,
\label{dimMorita}\end{equation} and hence that the canonical trace on
$\module$ is rescaled as \beq \Tr'=|r\,\theta+s|\,\Tr \ .
\label{TrMorita}\end{equation} In particular, the Morita equivalence
bimodule for the Heisenberg modules $\module_{1,0}$ and
$\module_{p,q}$ can be constructed by quantizing open strings in a
background $B$-field where the strings
stretch from a single D2-brane to a
cluster of $p$ coincident D2-branes carrying $q$ units of D0-brane
charge~\cite{sw}.

Under the gauge Morita equivalence the field strengths of
connections on the modules $\module$ and $\module'$ are related
through~\cite{gme}
\beq F'_{A'}=(r\,\theta+s)^2\,F_A+2\pi\,r\,(r\,\theta+s) \
. \label{FMorita}\end{equation} The invariance of the noncommutative
Yang-Mills action (\ref{YMactiondef}) is then guaranteed by the
transformation properties \bea g'^{\,2}&=&|r\,\theta+s|^3\,g^2 \ ,
\non\phi'&=& (r\,\theta+s)^2\,\phi-2\pi\,r\,(r\,\theta+s) \ .
\label{gphiMorita}\eea This mapping exhibits explicitly the
physical role \cite{sw,amns,gme,schwarzPhi,seiberg}
of the background flux $\phi$, as it is required to
absorb the inhomogeneous term in the transformation
(\ref{FMorita}). In particular, it can shifted in and out of the
action in a variety of different ways with no effect on the field
theory, thus showing precisely why it had no real bearing on our
previous analyses.

What is particularly important for us is the way that the solution
spaces of two dual gauge theories map into one
another~\cite{inprep}. The Morita transformation acts diagonally on
the critical point sets of noncommutative gauge theories, providing a
one-to-one correspondence between their constant curvature
connections~\cite{gme}, and hence their classical solutions. This
implies, in particular, that the basic structure of partitions between
dual gauge theories is the same. It is straightforward to check that
the classical action (\ref{NCYMpartition}) is invariant under the
transformations (\ref{thetaMorita}), (\ref{pqdoublet}) and (\ref{gphiMorita}).
Moreover, in the supersymmetric gauge theory the entire spectrum
of BPS states is invariant under the Morita duality~\cite{schwarzPhi}.

Finally, let us note that an ordinary $\theta=0$ gauge theory is
mapped to a noncommutative gauge theory with rational-valued
noncommutativity parameter $\theta'=n/s$ and Yang-Mills coupling
constant $g'^{\,2}=s^3\,g^2$ under the Morita map
(\ref{thetaMorita}) and (\ref{gphiMorita}). In this way it is
possible to embed ordinary Yang-Mills theory into the wider class
of noncommutative gauge theories, and this embedding has been used
to suggest that noncommutative Yang-Mills theory can be used to
impose very stringent constraints on ordinary large $N$ gauge
theories on tori~\cite{gt}. In particular, an ordinary $U(p)$ gauge theory
with 't~Hooft flux is Morita equivalent to a rational
$\theta'=n/s$ noncommutative gauge theory with single-valued gauge
fields~\cite{amns}. In this way the non-Abelian colour degrees of freedom of
commutative Yang-Mills theory may be traded for spacetime
noncommutativity.

\subsection*{Acknowledgments}

R.J.S. would like to thank the organisors and participants of the
meetings for the many questions and comments which have helped to
improve the material presented here, and also for the very
pleasant scientific and social atmospheres. The work of R.J.S. was
supported in part by an Advanced Fellowship from the Particle
Physics and Astronomy Research Council~(U.K.).

\end{document}